\documentstyle[12pt] {article}
\def\zid{1\kern-0.36em\llap~1}

\newcommand{\beq}{\begin{equation}}
\newcommand{\ber}{\begin{eqnarray}}
\newcommand{\eeq}{\end{equation}}
\newcommand{\eer}{\end{eqnarray}}

\begin{document}

\rightline{ SUNY BING 6/14/96}

\begin{center}
{\large \bf TESTS AT A TAU/CHARM FACTORY WITH LONGITUDINALLY
POLARIZED BEAMS}\\[2mm]
Charles A. Nelson\footnote{Electronic address: cnelson @
bingvmb.cc.binghamton.edu.  Contributed paper to ICHEP96,
Warsaw. } \\
{\it Department of Physics\\State University of New York at
Binghamton\\
Binghamton, N.Y. 13902-6016}\\[5mm]
\end{center}


\begin{abstract}

In a recent paper, eight semileptonic parameters were defined
to specify the most general Lorentz-invariant spin
correlation functions for tau semileptonic decays.  The
parameters were physically defined in terms of tau-decay
partial-width intensities for polarized final states.  This
paper studies how these parameters can be simply measured at
a tau/charm factory with longitudinally polarized beams
without using spin-correlation techniques.  Thereby the
parameters can also be used to bound the effective-mass
scales $ \Lambda $ for ``new physics" such as arising from
lepton compositeness, leptonic CP violation, leptonic T
violation, tau weak magnetism, weak electricity, and/or
second-class currents.
\end{abstract}

\newpage

\section{INTRODUCTION}

Currently, the bounds are very weak for possible ``new
physics" in tau lepton phenomena.  One of the ways in
which
more significant constraints could be obtained, or ``new
physics"
be discovered, would be through experiments at a tau/charm
factory \cite{a1,b1} which study the structure of the
$J^{charged}_{lepton}$ current.

In this paper we compare two simple methods for model-
independent determinations of the complete Lorentz structure
of tau semileptonic decays $\tau
^{-}\rightarrow \rho ^{-}\nu ,a_1^{-}\nu $. The first method
is to make use of longitudinally polarized electron and
positron beams (cf. Ref. 2) to study the decay chain $
 \tau ^{-}\rightarrow \rho ^{-}\nu
$\ followed by $\rho ^{ch}\rightarrow \pi
^{ch}\pi ^o$. The second method is
to make use of spin-correlations \cite{a2,c1,C94,pl,gt} to
study the decay sequence
$\gamma^{*}\rightarrow \tau ^{-}\tau ^{+}\rightarrow (\rho
^{-}\nu
)(\rho ^{+}\bar
\nu )$\ followed by $\rho ^{ch}\rightarrow \pi
^{ch}\pi ^o$.  In both methods we include both
$\nu _L$,\ $\nu _R\ $helicities and both   $\bar \nu _R$,\
$\bar \nu _L\
$helicities.  Similarly, we study \cite{ach2,pl,gt} the decay
chain $\tau^{-
}\rightarrow a_1^{-}\nu$ followed by $ a_1 \rightarrow
(3\pi)^{-} $.

In each case, we concentrate on a single
distribution function (in this paper) and determine the
associated ``ideal
statistical errors". \newline

{\bf $\cal{P}_L$ method}
For the case of longitudinally
polarized beams, we assume 100 \% polarization and study
the 3-variable distribution ${I_3}^{\cal P} ( \theta_{beam},
E_{\rho^-}, \tilde{\theta}_{\pi^{-}} )$.  In the center-of-
mass frame,  $\theta_{beam}$ is the angle between the final
tau momentum and the initial $e^-$ beam, and  $E_{\rho^-}$ is
the energy of the final $\rho^-$.  The angle
$\tilde{\theta}_{\pi^{-}}$ is the direction of the final
$\pi^-$ momentum in the $\rho^-$ rest frame [when boost is
directly from the center-of-mass frame]. \newline

{\bf Stage-two Spin-Correlation method}
We use the simple 4-variable distribution \cite{c1,pl,gt}
which includes information on $\rho ^{ch}\rightarrow \pi
^{ch}\pi ^o$.   It doesn't include information on the initial
$e^-$ direction.  This distribution is $I_4 (E_{\rho^-
},E_{\rho^+},\tilde{\theta}_{\pi^{-
}},\tilde{\theta}_{\pi^{+}} )$. \newline

{\bf Remarks:}
(1) The associated ideal statistical errors are given in the
tables at the end of this paper. \newline
(2) While more complicated distributions can be used, for a
preliminary comparison of the two methods these are
partcularly simple distributions which, nevertheless, provide
significant ``analyzing powers". \newline
(3) In the present paper, as previously \cite{c1,pl,gt}, we
assume a $10^7$ ($\tau ^{-}\tau ^{+}$%
) pair data sample at $4GeV$. We use branching ratios of 24.6
\% for
the $\rho$ mode, and 18 \% for the sum of the neutral and
charged $a_1$ modes, with an $a_1$ mass of 1.275 GeV.
\newline
(4)  Instead of $\theta_{beam}$ which is the angle between
the
$e^-$ and $\tau^-$ momenta, one could use the angle between
the $e^-$ and the final $\rho^-$ momenta.  This would not
require knowledge of the final tau direction.  Work on this
alternative 3-variable distribution is in progress \cite{ft}.

\section{SIMPLE 3-VARIABLE DISTRIBUTION IN CASE OF
LONGITUDINALLY POLARIZED BEAMS}

For the tau decay sequence $ \tau^{-}\rightarrow \rho^{-}\nu
$ followed by $\rho ^- \rightarrow \pi^- \pi ^o$,
\begin{equation}
I_3^{\cal P} ( \theta_B, E_{\rho^-},
\tilde{\theta}_{\pi^{-
}} ) = \sum \rho_{hh}^{prod} (e^- e^+ \rightarrow \tau^-
\tau^+ ) \rho_{hh}( \tau^- \rightarrow \rho^{-}\nu
\rightarrow \pi^- \pi ^o \nu )
\end{equation}
The summation is over the $\tau^-$ helicity, $h=\pm 1/2$.
Since the
azimuthal angle of the beam direction has been integrated
over, the off-diagonal elements in the density matrices do
not appear in Eq.(1).

The production density matrix for the $\tau^-$ is

\begin{equation}
\rho _{\lambda _1,\lambda _1^{^{\prime }}}^{LR}=\left(
\begin{array}{cc}
\sin ^4\theta _B /2+\frac{m^2}s\sin ^2\theta _B & -\frac m{
\sqrt{s}}e^{-\iota \Phi _B}\sin \theta _B \\ -\frac
m{\sqrt{s}}e^{\iota \Phi
_B}\sin \theta _B & \cos ^4\theta _B /2 +\frac{m^2}s\sin ^2
\theta _B
\end{array}
\right)
\end{equation}

\begin{equation}
\rho _{\lambda _1,\lambda _1^{^{\prime }}}^{RL}=\left( -
\right) ^{\lambda
_1-\lambda _1^{^{\prime }}}(\rho _{-\lambda _1,-\lambda
_1^{^{\prime
}}}^{LR})^{*}
\end{equation}
(i.e. these are used in the case when only $\tau^-$ decay
products are observed).  In these equations, $\theta
_B=\theta _{beam}$ is the angle between the
$e^{-}$and the $%
\tau ^{-}$momenta (and between the $e^{+}$ and the
$\tau ^{+}$ momenta).

For the antiparticle leg with the $\tau^+$ but referred still
to the $e^-$ beam, the production density matrix  is
\begin{equation}
\bar \rho _{\lambda _2,\lambda _2^{^{\prime }}}^{LR}=\left(
\begin{array}{cc}
\cos ^4\theta _B /2+\frac{m^2}s\sin ^2\theta _B & -\frac m{
\sqrt{s}}e^{\iota \Phi _B}\sin \theta _B \\ -\frac
m{\sqrt{s}}e^{-\iota \Phi
_B}\sin \theta _B & \sin ^4\theta _B /2 +\frac{m^2}s\sin ^2
\theta _B
\end{array}
\right)
\end{equation}
\begin{equation}
\bar \rho _{\lambda _2,\lambda _2^{^{\prime }}}^{RL}=\left( -
\right)
^{\lambda _2-\lambda _2^{^{\prime }}}(\bar \rho _{-\lambda _2,
-\lambda
_2^{^{\prime }}}^{LR})^{*}
\end{equation}
(i.e. these are used in the
case when only $\tau^+$ decay products are observed).
products are observed).  In Eq.(5), $\theta
_B$ is still the angle between the
$e^{-}$and the $%
\tau ^{-}$momenta (and between the $e^{+}$ and the
$\tau ^{+}$ momenta).

\section{STAGE-TWO SPIN-CORRELATION FUNCTION $I_4$}

For comparison with the case of
longitudinally-polarized beams, we use the simple 4-variable
S2SC function

\begin{equation}
\begin{array}{c}
{I(E}_\rho {,E}_{\bar \rho }{,}\tilde \theta
_1{,}\tilde
\theta _{2{ }}{) = }|{T}\left( +-\right)
|^2\rho
_{++}\bar
\rho _{--} +{ }|{T}\left( -+\right) |^2\rho _{--
}\bar \rho
_{++} \nonumber  \\ +{ }|{T}\left( ++\right)
|^2\rho
_{++}\bar \rho
_{++}
 +{ }|{T}\left( --\right) |^2\rho
_{--}\bar \rho _{--}
\end{array}
\end{equation}
In terms of probabilities, the quantum-mechanical
structure of this
expression is apparent,  since the $T(\lambda _{\tau
^{-}},\lambda _{\tau ^{+}})$ helicity amplitudes
describe the production of the $(\tau ^{-}\tau ^{+})$ pair
via $Z^o$,
or $%
\gamma ^{*}\rightarrow \tau ^{-}\tau ^{+}$. For instance, in
the 1st
term,
the factor  $|T(+,-)|^2=$``Probability to produce a $\tau ^{-
}$ with
$%
\lambda _{\tau ^{-}}=\frac 12$ and a $\tau ^{+}$ with
$\lambda
_{\tau
^{+}}=-\frac 12$ '' is multiplied by the product of the decay
probablity, $\rho _{++}$, for the positive helicity $\tau ^{-
}\rightarrow
\rho ^{-}\nu \rightarrow \left( \pi ^{-}\pi ^o\right) \nu $
times the
decay
probablity, $\bar \rho _{--}$, for the negative helicity
$\tau
^{+}\rightarrow \rho ^{+}\bar \nu \rightarrow \left( \pi
^{+}\pi
^o\right)
\bar \nu $ .

\section{COMPARISON OF TWO METHODS}

Both of the above methods involve the same composite decay
density matrices for $\tau ^{-
}\rightarrow
\rho ^{-}\nu \rightarrow \left( \pi ^{-}\pi ^o\right) \nu $,
\dots, and similarly for the $a_1$ decay mode.  So when
defining the parametrization of
these decay matrices, it is
convenient to simultaneously report the associated "ideal
statistical errors" .

{\bf I: Measurement of general semileptonic parameters:}

The 8 tau semi-leptonic decay parameters \cite{gt} for $\tau
^{-
}\rightarrow \rho
^{-}\nu, \ldots $, are defined for the four polarized
$\rho_{L,T} \nu_{L,R}
$ final states:  The first parameter is simply $\Gamma \equiv
\Gamma
_L^{+}+\Gamma _T^{+}$, i.e. the partial width
for $\tau ^{-}\rightarrow \rho ^{-}\nu $.  The second is the
chirality
parameter $ \xi \equiv \frac 1\Gamma (\Gamma _L^{-}+\Gamma
_T^{-}) $. Equivalently,
\newline $ \xi \equiv$ (Prob $\nu_{\tau}$ is
$\nu_L$) $ - $
(Prob $\nu_{\tau}$ is $\nu_R$), or
\begin{equation}
\xi \equiv |< \nu_L |\nu_{\tau} >|^{2} - |< \nu_R |\nu_{\tau}
>|^{2}
\end{equation}
So a value $\xi = 1$ means the coupled $\nu_{\tau}$ is
pure $\nu_L$.  $\nu_L$ ($\nu_R$) means the emitted neutrino
has
L-handed (R-handed) polarization.
For the special case of a mixture of only $V$ \& $A$
couplings and $m_{
\nu_{\tau} } = 0 $, $\xi \rightarrow \frac{\left| g_L\right|
^2-\left|
g_R\right| ^2}{\left| g_L\right| ^2+\left|g_R\right| ^2}$ and
the ``stage-one spin correlation" parameter $\zeta
\rightarrow \xi$.  The subscripts on the $\Gamma $'s denote
the polarization of the final $\rho ^{-}$, either
``L=longitudinal'' or
``T=transverse''; superscripts denote ``$\pm $ for
sum/difference of
the $\nu _{L\ }$versus $\nu _R$ contributions''.

The remaining partial-width parameters are defined by
\begin{equation}
\zeta \equiv (\Gamma _L^{-}-\Gamma _T^{-})/(
{\cal S}_\rho \Gamma ), \hspace{2pc} \sigma \equiv (\Gamma
_L^{+}-\Gamma
_T^{+})/(
{\cal S}_\rho \Gamma ).
\end{equation}
To describe the interference between the $\rho_L$ and
$\rho_R$ amplitudes,
we define
\begin{equation}
\begin{array}{c}
\omega \equiv I_{
{\cal R}}^{-}\ /({\cal R}_\rho \Gamma ), \hspace{2pc}  \eta
\equiv I_{
{\cal R}}^{+}\ /({\cal R}_\rho \Gamma ) \\ \omega ^{\prime
}\equiv I_{
{\cal I}}^{-}\ /({\cal R}_\rho \Gamma ), \hspace{2pc} \eta
^{\prime }\equiv
I_{{\cal I}%
}^{+}\ /({\cal R}_\rho \Gamma )
\end{array}
\end{equation}
where the measureable $LT$-interference intensities are
\begin{equation}
\begin{array}{c}
I_{{\cal R}}^{\pm }=\left| A(0,-\frac 12)\right| \left| A(-
1,-\frac
12)\right| \cos \beta _a \pm \left| A(0,\frac 12)\right|
\left| A(1,\frac
12)\right| \cos \beta _a^R  \\
I_{{\cal I}}^{\pm }=\left| A(0,-\frac 12)\right| \left| A(-
1,-\frac
12)\right| \sin \beta _a \pm \left| A(0,\frac 12)\right|
\left| A(1,\frac
12)\right| \sin \beta _a^R
\end{array}
\end{equation}
Here $\beta _a\equiv \phi _{-1}^a-\phi _0^a$, and $\beta
_a^R\equiv \phi
_1^a-\phi _0^{aR}$\ are the measurable phase differences of
of the
associated helicity amplitudes
$A(\lambda_{\rho},\lambda_{\nu})=\left|
A\right| \exp \iota \phi $.

Four of these parameters ($\xi, \zeta, \sigma, \omega$)
appear in the $\rho_{hh}$ density
matrix which occurs in the above distribution functions,
$I_3^{\cal{P}}$ and $I_4$.

{\bf Formulas for }$\tau \rightarrow \rho \nu :$

The composite decay density matrix elements are
simply the
decay
probability for a $\tau _1^{-}$ with
helicity $\frac h2$ to decay $\tau ^{-}\rightarrow \rho ^{-
}\nu
\rightarrow
\left( \pi ^{-}\pi ^o\right) \nu $ since
\beq
\frac 1{\Gamma } \frac{d{N}}{d\left( \cos \theta _1^\tau
\right) d\left(
\cos \tilde
\theta _1\right) }=\rho _{hh}\left( \theta _1^\tau ,\tilde
\theta
_1\right)
\eeq

\noindent
and for the decay of the $\tau _2^{+}$ ,
\beq
\bar \rho _{hh}=\rho _{-h,-h}\left( {subscripts} \quad
1
\rightarrow 2, a
\rightarrow
b \right)
\eeq
For a $\tau _1^{-}$ with
helicity $\frac h2$ to decay $\tau ^{-}\rightarrow \rho ^{-
}\nu
\rightarrow
\left( \pi ^{-}\pi ^o\right) \nu $,
\ber
\begin{array}{c}
{\bf \rho }_{hh} =\frac 1{8}(3+\cos
2\tilde \theta
_1)S+\frac 1{32}(1+3\cos 2\tilde \theta _1)D
\end{array}
\eer
where
\beq
S=1+h\zeta {\cal S}_\rho  \cos \theta _1^\tau
\eeq
\beq
D=-S(1-\cos 2\omega _1)+(\sigma {\cal S}_\rho +h\xi \cos
\theta _1^\tau )(1+3\cos 2\omega
_1)+h\omega {\cal R}_\rho 4\sqrt{2}\sin 2\omega _1\sin \theta
_1^\tau .
\eeq
with the Wigner rotation angle $\omega_1 =
\omega_1(E_{\rho})$, \cite{C94}.

{\bf Formulas for }$\tau \rightarrow a_1\nu :$

For $\tau ^{-}\rightarrow a_1^{-}\nu \rightarrow (3\pi )^{-
}\nu $, with $%
\tau ^{-}$\ helicity $\lambda _1=h/2$ where%
\ber
\begin{array}{c}
{\bf \rho }_{hh} =\frac 14(3+\cos
2\tilde \theta
_1)S_{a_1}-\frac 1{32}(1+3\cos 2\tilde \theta _1)D_{a_1}
\end{array}
\eer
\beq
S_{a_1}=1+h\zeta {\cal S}_{a_1} \cos \theta _1^\tau
\eeq
\beq
D_{a_1}=S_{a_1}(3+\cos 2\omega _1)+(\sigma {\cal S}_{a_1}
+h\xi \cos \theta _1^\tau
)(1+3\cos 2\omega _1)+h\omega {\cal R}_{a_1} 4\sqrt{2}\sin
2\omega _1\sin \theta _1^\tau .
\eeq
For the CP conjugate process, $\tau ^{+}\rightarrow
a_1^{+}\bar \nu
\rightarrow (3\pi )^{+}\bar \nu $, with $\tau ^{+}$\ helicity
$\lambda _2=h/2$,
\beq
\bar \rho _{hh}=\rho _{-h,-h}\left( {subscripts} \quad
1
\rightarrow 2, a
\rightarrow
b \right)
\eeq

{\bf Ideal statistical errors for measurement of $\xi, \zeta,
\sigma,$ and $\omega$:}

For the $10^7$ ($\tau^-,\tau^+$)'s at 4 GeV, we determine the
ideal statistical errors in the same manner as in our earlier
papers, see Ref. 4.

See Table 1 for the errors for ($\xi, \zeta, \sigma, \omega$)
based on $I_3^{\cal{P}}$ and on $I_4$. In general, by using
longitudinally-polarized beams the errors for the $\rho^-$
mode are slightly less than $ 0.4 \% $ and about a factor of
7 better than by using the S2SC function $I_4$.  The
CP tests for these semileptonic parameters are $\sqrt2$
worse by the $\cal{P}_L$ method, and about the same by the
S2SC method.  Typically the $a_1$ values are 2-4 times worse
than the $\rho$ values. However, for $\xi$, the error for the
$a_1$ mode by the $\cal{P}_L$ method is about 3 times better
than that for the $\rho$ mode.

{\bf II:  Two tests for non-CKM-type leptonic $CP$ violation
if only $\nu_L$ and $\bar{\nu}_R$ couplings:}

Here we use a different parametrization of the composite
decay density matrix since we assume only $\nu _L$
couplings.  For the $\rho$ mode we use \cite{C94}
\ber
\rho _{hh}=
  \left( 1+h\cos \theta _1^\tau \right) \left[ \cos ^2\omega
_1\cos^2\tilde \theta
_1+\frac 12\sin ^2\omega _1\sin ^2\tilde \theta _1\right]
\nonumber
\\
+ \frac{r_a^2}2\left( 1-h\cos \theta _1^\tau \right)
\left[ \sin ^2\omega_1\cos ^2\tilde \theta _1   +\frac
12\left( 1+\cos
^2\omega
_1\right) \sin^2\tilde \theta _1 \right]   \nonumber \\
 +h\frac{r_a}{\sqrt{2}}\cos \beta _a\sin \theta
_1^\tau \sin 2\omega _1\left[ \cos ^2\tilde \theta _1-\frac
12\sin
^2\tilde
\theta _1\right]
\eer
The dynamical parameters to be experimentally measured are
the polar parameters
$\beta _a=\phi
_{-1}^a-
\phi _0^a$,
$\beta
_b=\phi _1^b-\phi _0^b$, and $r_a={|A\left( -1,-\frac
12\right) |}/{|A\left( 0,-\frac
12\right) |}$, $r_b={|B\left( 1,\frac 12\right) |}/{|B\left(
0,\frac 12\right) |}$.  In the
standard lepton model with a pure $(V-A)$ coupling, the
predicted
values are $\beta _{a,b}=0,r_{a,b}=\frac{\sqrt{2}m_\rho
}{E_\rho
+q_\rho }\simeq \sqrt{2}m_\rho /m_\tau \simeq 0.613.$

For  the $\tau ^{-}\rightarrow
a_1^{-}\nu \rightarrow \left( \pi ^{-}\pi
^{-}\pi ^{+}\right) \nu ,\left( \pi ^o\pi ^o\pi ^{-}\right)
\nu $ modes,
\ber
\rho _{hh}=
 \left( 1+h\cos \theta _1^\tau \right) \left[ \sin ^2\omega
_1\cos
^2\tilde \theta _1
+ ( 1- \frac 12\sin ^2\omega _1 ) \sin ^2\tilde \theta
_1\right]
\nonumber \\
+ \frac{r_a^2}2\left( 1-h\cos \theta _1^\tau \right)   \left[
\left(
1+\cos
^2\omega
_1\right) \cos ^2\tilde \theta _1  +\left( 1+\frac 12\sin
^2\omega
_1\right) \sin
^2\tilde \theta _1 \right]  \nonumber  \\
-h\frac{r_a}{\sqrt{2}}\cos \beta _a\sin \theta
_1^\tau \sin 2\omega _1\left[ \cos ^2\tilde \theta _1-\frac
12\sin
^2\tilde
\theta _1\right]
\eer
Here $\tilde \theta _1$ specifies the normal to the $\left(
\pi ^{-}\pi
^{-}\pi ^{+}\right) $ decay triangle, instead of the $\pi ^{-
}$
momentum
direction used for $\tau ^{-}\rightarrow \rho ^{-}\nu $.  The
Dalitz
plot for $\left( \pi ^{-}\pi ^{-}\pi ^{+}\right) $ has been
integrated
over
so that it is not necessary to separate the
form-factors for $a_1^{-} \rightarrow $ $\left( \pi ^{-}\pi
^{-}\pi
^{+}\right) $.  In the
standard lepton model with a pure $(V-A)$ coupling, for the
$a_1$ mode $r_{a,b}= 1.01$ for $m_{a_1} = 1.275 GeV$.

{\bf Ideal statistical errors for two tests for "non-CKM-
type" leptonic CP violation:}

Tables 2 \& 3 show respectively the sensitivities of the
$\rho$ and $a_1$ modes for measurements by the two tau-
polarization methods.  By either polarization technique, the
moduli ratio $r_a$ versus $r_b$ can be measured to better
than $0.1\%$.  The phase differences $\beta_a, \beta_b$ can
be measured to about $7^o$ by these techniques; however, in
the S2SC case the $I_7$ distribution is about 2 times as
sensitive so inclusion of more variables to describe the
final state may also give significant
improvement in the case of longitudinally-polarized beams.

{\bf III: Measurement of effective-mass scales $\Lambda$ for
additional ``Chiral Couplings":}

In Ref. 7, the above semileptonic parameters have been
expressed interms of additional ``chiral couplings" in the
charged-current which could arise due to ``new physics".

The most general Lorentz coupling for \hskip
1em  $\tau^{-
}\rightarrow \rho
^{-}\nu _{L,R}$ is
\beq
\rho _\mu ^{*}\bar u_{\nu _\tau }\left( p\right) \Gamma ^\mu
u_\tau
\left(
k\right)
\eeq
where $k_\tau =q_\rho +p_\nu $. It is convenient to treat the
vector
and
axial vector matrix elements separately. In Eq.(22)
$$
\Gamma _V^\mu =g_V\gamma ^\mu +
\frac{f_M}{2\Lambda }\iota \sigma ^{\mu \nu }(k-p)_\nu   +
\frac{g_{S^{-}}}{2\Lambda }(k-p)^\mu +\frac{g_S}{2\Lambda
}(k+p)^\mu
+%
\frac{g_{T^{+}}}{2\Lambda }\iota \sigma ^{\mu \nu }(k+p)_\nu
$$
\beq
\Gamma _A^\mu =g_A\gamma ^\mu \gamma _5+
\frac{f_E}{2\Lambda }\iota \sigma ^{\mu \nu }(k-p)_\nu \gamma
_5
+
\frac{g_{P^{-}}}{2\Lambda }(k-p)^\mu \gamma
_5+\frac{g_P}{2\Lambda }%
(k+p)^\mu \gamma _5  +\frac{g_{T_5^{+}}}{2\Lambda }\iota
\sigma ^{\mu \nu
}(k+p)_\nu \gamma _5
\eeq

The parameter
$%
\Lambda =$ ``the effective-mass scale of New Physics''. In
effective field
theory
this
is the scale at which new particle thresholds are expected to
occur or where the theory becomes non-perturbatively
strongly-interacting so as to overcome perturbative
inconsistencies.  It can also be interpreted as a measure of
a
new compositeness scale.  In old-fashioned renormalization
theory
$\Lambda$  is the scale at
which the calculational methods and/or the principles of
``renormalization''
breakdown. Without additional theoretical or
experimental
inputs, it is not possible to select what is the "best"
minimal set of couplings for
analyzing the structure of the tau's charged current.  For
instance, by  Lorentz
invariance, there are the equivalence theorems that for the
vector
current%
\ber
S\approx V+f_M, & T^{+}\approx -V+S^{-}
\eer
\noindent
and for the axial-vector current
\ber
P\approx -A+f_E, & T_5^{+}\approx A+P^{-}
\eer
On the other hand, dynamical considerations such as lepton
compositeness
would suggest searching for an additional tensorial
$g_{+}=f_M +
f_E$ coupling which would preserve $\xi =1$ but otherwise
give
non-($V-A$)-values to the semi-leptonic parameters. For
instance, $\sigma =\zeta
\neq 1$and $%
\eta =\omega \neq 1$.

{\bf Effective-mass scale bounds for additional ``chiral
couplings":}

Tables 4 \& 5 respectively give the limits \cite{pl} on
$\Lambda$ in the
case of purely real and imaginary coupling constants for
additional ``chiral couplings".  Scales of the order of $1
TeV$ can be probed for some real coupling constants.

We list the ideal
statistical error for the presence of an additional $V+A$
coupling as an error $\delta (\xi _A)$ on the chirality
parameter $\xi _A$
for $\tau ^{-}\rightarrow A^{-}\nu $. Equivalently, if one
ignores possible
different L and R leptonic CKM factors, the effective lower
bound on an
additional $W_R^{\pm }$ boson (which couples only to right-
handed
currents)
is
\ber
M_R=\{\delta (\xi _A)/2\}^{-1/4}M_L
\eer
So $\delta (\xi)=0.0012(0.0018)$ respectively correspond to
$M_R>514GeV(464GeV)$.

In some cases for real coupling constants, the S2SC method
gives a bound about a factor of 2 better than that for the
$I_3^{\cal{P}}$ method.  Here also it is important to extend
the present analysis in the case of longitudinally-polarized
beams to see what occurs when addtitional
variables are included in the description of the final state.

\section{CONCLUSIONS}

In the present paper two simple tau-polarization techniques
have been compared for possible use at a tau/charm factory to
study the $J_{lepton}^{charged}$ current.  For measurement of
the semileptonic parameters, $\xi, \zeta, \sigma, \omega$,
the $\cal{P}_L$ method using $I_3^{\cal{P}}$ is about 7 times
better than the S2SC($I_4$) method.  Both methods are
comparable for the two tests for non-CKM-type leptonic $CP$
violation.  In some cases the S2SC($I_4$) method gives about
a 2 times stronger bound on addtional chiral couplings.

In the case of the S2SC method, additional kinematic
variables have been shown to be important to include in
describing the final state.  Thereby additional semileptonic
parameters can be measured and significantly greater
analyzing powers can be achieved.  More analysis is needed to
see if the same is true when additional variables are
included in the case of the $\cal{P}_L$ method which exploits
longitudinally-polarized beams. This should be true because
the 3 variables so-far included in $I_3^{\cal{P}}$ do not
fully exploit the special kinematics of the tau threshold
region.

\section*{Acknowledgments}
For helpful discussions, we thank experimentalists and
theorists involved in on-going and potential tau lepton
experiments.  This work was partially supported by U.S. Dept.
of Energy Contract No. DE-FG 02-96ER40291.

\section*{Table Captions}
\quad Table 1: Ideal statistical errors for measurements at 4
GeV of the fundamental parameters $ \xi, \zeta,
\sigma$, and $ \omega$ by either (i) the simple $
I_3^{\cal{P}}$ distribution function for $
\tau^- \rightarrow \rho^- \nu $ using longitudinally-
polarized $e^- e^+$ beams, or by (ii) the stage-two spin-
correlation
function $ I_4 $ for the sequential decay of an off-mass-
shell photon $ \gamma^* \rightarrow \tau^- \tau^+ $ with $
\tau^- \rightarrow \rho^- \nu $ and $ \tau^+ \rightarrow
\rho^+ \bar\nu $, etc.
For each parameter, the first row assumes CP-invariance, for
instance $ \xi = \bar\xi $; then the following row contains
the corresponding the statistical errors for measurement of
the same parameter not assuming CP-invariance.  We use $10^7$
$ \gamma^* \rightarrow \tau^- \tau^+ $ events.

Table 2: Ideal statistical errors for two tests for $CP$
violation in $\tau \rightarrow \rho \nu $ by (i) the S2SC
function, $I_4$, or by (ii) the longitudinally-polarized beam
distribution function, $I_3^{\cal{P}}$.

Table 3: Ideal statistical errors for two tests for $CP$
violation in $\tau \rightarrow a_1 \nu $ by (i) the S2SC
function, $I_4$, or by (ii) the longitudinally-polarized beam
distribution function, $I_3^{\cal{P}}$.

Table 4: ``Chiral Couplings'':  Limits on $\Lambda$ in $GeV$
for
real coupling constants. For the $\rho$ and $a_1$ modes, the
$T^{+}+T_5^{+}$ coupling is equivalent to the $V-A$ coupling;
and $T^{+}-
T_5^{+}$ is equivalent to $V+A$.   For $V+A$ only, the entry
is for
$\xi_A$;  by Eq.(26) these values can be converted to a bound
on the $M_R$ mass of a R-handed $W^{\pm}$.

Table 5: ``Chiral Couplings'':  Limits on $\Lambda$ in $GeV$
for
pure imaginary coupling constants. For the $\rho$ and $a_1$
modes, the
$T^{+}+T_5^{+}$ coupling is equivalent to the $V-A$ coupling;
and $T^{+}-
T_5^{+}$  is equivalent to the $V+A$.

{\bf Tables are available by airmail or FAX---contact author by
email.}

\end{document}